\magnification 1200
\centerline{\bf A program for a problem free Cosmology within a}
\centerline{\bf framework of a rich class of scalar tensor theories}
\vskip 3cm
\centerline {Daksh Lohiya \& Meetu Sethi}
\centerline{Department of Physics and Astrophysics, University of
Delhi}
\centerline{Delhi 110 007, India}
\centerline {email: dlohiya@ducos.ernet.in; meetu@ducos.ernet.in}

\vskip 1cm
\centerline{\bf Abstract}
     
     A search for a problem free cosmology within the framework
of an effective non - minimally coupled scalar tensor theory 
is suggested. With appropriate choice of couplings in variants of 
a Lee - Wick model [as also in a model supporting Q - ball solutions],
non topological solutions [NTS's], varying in size upto 10 kpc to 1
Mpc can exist. We explore the properties of a ``toy'' Milne model 
containing a distribution of NTS domains. 
The interior of these domains
would be regions where effective gravitational effects would be 
indistinguishable from those expected in standard Einstein theory.
For a large class of non - minimal coupling terms and the scalar 
effective potential, the effective cosmological constant
identically vanishes. The model passes classical
cosmological tests and we describe reasons to expect it 
to fare well as regards nucleosynthesis and structure formation.

\vfil\eject

\centerline{\bf I. Introduction}
\vskip 1cm

      A description of a Friedman - Robertson - Walker [FRW]
universe within the framework of general theory of relativity [GTR]
has its own peculiar problems. Besides having an initial
singularity,
a lack of a consistent quantum mechanical framework and a lack of 
a consistent account of large scale structure formation in the
theory,
the observed large scale homogeneity and isotropy can not be 
dynamically generated in standard GTR due to the so called 
``horizon problem''. 
The stability of the FRW solution, moreover, requires a fine 
tuning of the density parameter in the theory unless it is exactly
equal to the 
critical density. This is referred to as the flatness problem. 
A resolution of these problems using inflation has developed into 
a state of art over the last almost two decades. During the
inflationary epoch, the energy density of the universe is dominated by
the vacuum energy of an inflaton field
while the scale factor expands superluminally. With
a sufficiently large interval of exponential inflation, a small 
causally connected region of the universe grows sufficiently large to 
account for the the observed homogeneity and isotropy of the universe.
In most versions, however, this paradigm requires a very special profile 
for the effective potential for scalar field [s] in the model on
account of constraints on inflation [1]. The requirement
of sufficient
inflation and CMB anisotropy limits  the density
fluctuations [2] and 
in turn constrains the inflaton field potential to 
be very flat. For a general class of inflationary models involving a 
slow rolling field [including new [3], chaotic [4] and multiple 
field [5] inflation], one needs the inflaton potential to
satisfy: $\Delta V/(\Delta\phi)^4 \leq 10^{-6} - 10^{-8}$ [6]. Where
$\Delta V$ and $\Delta\phi$ are changes in the potential and the field
during the slow rolling. For a quartic coupling
[$V\approx\lambda\phi^4$] for example, this presents a constraint 
$\lambda \leq 10^{-6} - 10^{-8}$ on the coupling constant. Small 
couplings at tree level are unnatural in general as they require 
fine tuning to cancell large radiative corrections. Standard 
convexity and triviality theorems in quantum field theory also cast a
doubt on whether
such an effective potential can at all be realised. Most versions 
of an appropriate inflation are further heavily constrained by baryon 
asymmetry constraints [the graceful exit and the reheating problems
(see eg.[7] for a recent review)].

     Assuming that these formal 
problems would be resolved, inflationary scenarios ensure the
favoured value for the
density parameter of the FRW universe viz.: the closure density, in
turn ensuring 
the density parameter
$\Omega = 1$. This defines what is now regarded as the so called
Standard - Big - Bang model [SBB]. Over the last decade, the SBB
has suffered
a lot of stress even as far as the post - inflationary empirical
observations are concerned. There has been a steady growth of  
evidence
indicating that $\Omega$ over scales of a giga parsec and more is 
significantly less than one. Figures of .2 to .8 have been quoted in 
literature. The flatness problem [fine tuning] then stares
SBB in the face. This is not withstanding models [8] contrived 
to yield $\Omega$ less than unity as a result of multi - inflating 
epochs that judiciously combine an epoch of old inflation followed by 
a new - inflation wherein $\Omega$'s approach to unity is cut short 
arbitrarily. These models include ``extended and hyperextended''
inflationary models [9], 
in which both the inflaton and the Brans -  Dicke field come
with their respective potentials and a proliferation of parameters.
With better instrumentation and 
observations, the parameter space of the post inflationary SBB
universe is shrinking. The
worst constraints come from age estimates of old clusters
in comparison with age estimates of the universe from the measurement 
of the Hubble parameter. There is further evidence of excessive baryonic 
dark matter, inconsistent with SBB nucleosynthesis, 
from the intensity of x-ray from centres of
clusters of galaxies. Recent magnitude - redshift relations based on
type SN1A observations [10], rule out the $\Omega\geq 1$
models. It has been proposed that a small
cosmological constant be incorporated in the model. While this
would not do away with the fine tuning problem, the resolution that
it may achieve is seriously in doubt.
The SBB may well be on the verge of a crisis.

    To summarise: though inflation can resolve the horizon, 
flatness and the monopole problems and has a promise to give 
an ansatz for primordial density fluctuations, it is still too
early and naive to defend this paradigm as a faith. For one, the
inflaton field requires a fine tuning of its dynamical parameters
leaving hardly any ``naturalness'' in the model. Further,
a transition from an almost empty inflated patch to a matter
filled universe (the reheating and the graceful exit problems) 
can not be dynamically
realised in most versions of the inflationary 
scenario. Most versions of 
inflation find the smallness of the cosmological constant an 
embarrassment. The ``natural'' value of the cosmological
constant in most unification schemes, the inverse square
of the Planck length, differs from cosmological
observational bounds by some 120 orders of magnitude !
In its most acceptable form, it would be fair to view 
inflation as a paradigm that relates large scale properties
[over distances greater than the the hubble length scale at the 
last scattering surface] to the physics at Planck energy scale -
either through the peculiarities of a finely tuned [contrived]
scalar potential or through quantum gravity. As one hardly has any
testable attributes of Planck physics on laboratory scales,
there is a feeling of 
arbitrariness and little predictive power [11] 
in the exercise of fitting any cosmological
attribute to a corresponding attribute of the inflaton effective 
potential. 

         Features of inflation have been extended to scalar 
tensor theories in general. These include the Brans - Dicke scalar 
tensor theory together with a separate inflaton field mentioned 
earlier. Recent interest in these theories motivates from 
string theory. It is believed that so far the only consistent 
theory of quantum gravity is string theory [12]. 
The low energy effective action that follows 
from string theory has the form of a scalar tensor theory of
gravity with non - trivial coupling of the dilaton to matter
[13]. The entire excersize of constraining the 
extra parameters of the Brans - Dicke (dialaton) field, with 
or without an further parameters of an inflaton field, has been 
extensively described in literature [14]. Rather than 
looking for empirical fits in the framework of proliferating
parameters, it would be a relief if basic cosmic attributes
like the absence of the horizon and the fine tuning problems 
were to be explained for a general equation of state of matter
(including any ``dialaton'' or an ``inflaton'' fields). It is
with this hope that we proceed to describe features of a toy
model in the following:

        A dynamic realisation of large scale homogeneity and 
isotropy in a FRW universe for an arbitrary equation of state
requires the scale
factor to evolve as $t^n$ with $n \geq 1$. It turns out that
within the framework of a general scalar tensor theory and for 
a general equation of state of matter, the only possible value
is $n = 1$ [Milne model[15]]. As we shall see in sec III,
such a coasting cosmology is not excluded
by classical cosmological tests [16]. Such a 
scaling, however, can only be possible in general 
in a model in which long 
distance gravity vanishes. [There are special solutions in Brans -
Dicke theory that can also support such a scaling [17]. 
However: (a) these do not stand upto cosmological constraints 
and (b) these are not stable to perturbations of the 
parameters of the theory].  The most crucial requirement
for such an idea would be to specify
an ansatz that could account for a spatial variation of the 
gravitational constant that vanishes outside compact domains 
and be uniform in the interior. The ansatz ought to 
generate effective gravitating domains upto say 1MPc in size. 

     In this article we outline an effective 
gravity model of a scalar tensor [ST] theory
of gravity
in which a Higgs field $\phi$ is also
coupled to the scalar curvature $R$ through 
a function $U(\phi)$ which diverges at some point that we take as $\phi = 0$ 
Thus: $U(\phi\longrightarrow 0) 
\longrightarrow \infty$. With the Higgs generating masses for
Fermions, 
non - topological soliton solutions [NTS's]
arise in such theories [18]. We show that each NTS 
is a domain that separates the exterior region, 
where the non minimal coupling with the Ricci scalar
is chosen to diverge, from the interior where the scalar field could be 
arrested to an arbitrary value - depending on the profile of the effective
potential and the non - minimal coupling. The interior is thus a region 
with an effective gravitational constant and an effective 
cosmological constant determined by the interior values of 
the non minimal coupling term $U(\phi)$, and the
scalar effective potential $V(\phi)$ respectively.
The essential point is that we expect the scalar field in the
interior of larger and larger stable NTS domains to approach a value that has 
smaller and smaller $\mid V(\phi)\mid$.
In this limit,
the effective gravitational constant would approach a universal
value inside all large NTS's - determined by a value $\phi_{in}^o$ where 
the effective potential vanishes.
This would be an effective solution to the cosmological
constant problem. The effective cosmological constant would vanish
outside a NTS (as the effective gravitational constant is zero there)
and be near zero in the interior of all large NTS's. 
Similar solutions are also possible in a theory in which the scalar
curvature non - minimally couples to an invariant 
function of a  multi - component 
scalar field. Such possibilities have been explored recently
in a search of alternative inflationary models [19]. If the 
invariant NMC again diverges at 
$\mid\phi\mid \equiv \sqrt{\phi_1^2 + \phi_2^2} =0$
for [say] an SO(2) invariant field non - topological 
soliton solutions [NTS's] again exist [Q - balls] with properties
similar to those described above for the analogous Lee-Wick constructs.
The interior is again a
region 
with an effective gravitational constant and an effective 
cosmological constant determined by the interior values of 
the non minimal coupling term $U(\mid\phi\mid)$, and the
scalar effective potential $V(\mid\phi\mid)$ respectively.
We expect each NTS domain to expand
while conserving the charge and the energy of the NTS.
This necessarily implies that the expansion would be accompanied 
by a drift of $\mid\phi\mid$ to a value $\mid\phi_o\mid$ where
$V(\mid\phi\mid)$ vanishes. As $\mid\phi\mid \longrightarrow
\mid\phi_o\mid$,
the volume of the domain becomes large. In this limit,
the effective gravitational constant would approach a universal
value inside all NTS's - determined by a value $\mid\phi_0\mid$
where 
the effective potential vanishes.

	 The program essentially requires non - minimal
coupling [NMC]. There is no compelling principle to constrain the
coupling of a function of a scalar field with the scalar curvature.
Conformal coupling for 
a single component scalar field: $U(\phi) = \phi^2/6$ is known [20]
to give decent
renormalisable properties of the stress energy tensor. Zee [21] and  
earlier, Deser [22] have considered NMC's to generate
effective Brans - Dicke like theories virtually indistinguishable 
from general theory of relativity [GTR] 
at low energies. Dolgov [23] has 
explored [though not successfully] a rising NMC
as a mechanism to dynamically reduce the effective 
cosmological constant. Others [19] have used invariant NMC's in 
multicomponent extended inflationary models.
Madsen [24] 
has extensively reviewed properties of a large class of NMCS. 
There seems to be no consensus on any particular principle that
may be used to specify the NMC.
For our purpose we propose an ansatz that 
leaves the form for the NMC unspecified as an
otherwise arbitrary function $U(\phi)$ that diverges at $\phi = 0$
and has a sufficiently large gradient in an open interval
containing $\phi = \phi_{in}^o$ where the effective potential vanishes.
We require the model to support NTS's. This is what
is outlined in the next section where we describe constraints on a 
scalar field theory
in order that it generate an effective theory of gravitation 
indistinguishable from GTR. In the class of ST theories 
considered, we establish
the existence of solutions across the boundary of the NTS - connecting to
the flat spacetime in the exterior of the NTS. 

        Section III, 
translates some of the aspects of standard model in the 
framework of the ``toy'' model described here and
summarises the 
predictions of the model. In the appendix
we describe the essential properties of a 
scalar tensor theory. Expressions for a conserved 
energy - momentum pseudo - tensor are derived for a general non -
minimally 
coupled ST theory.
\vskip 2cm
\centerline{\bf II. A Milne model with a difference:}
\vskip 1cm
        We revisit the
favourite model of Milne [15], who considered the evolution of 
an exploding universe from a highly correlated state [a 
vanishingly small ball of particles of infinite density] 
localised near the neighbourhood of the point
x = y = z = 0 at t = 0, in a Minkowski spacetime. At any later 
time, the universe consists of an isotropically expanding 
swarm of particles of all speeds bounded by the speed of 
light. Special relativity tells us that the universe looks 
the same (isotropically expanding ball bounded by a light 
front) in all Lorentz frames that coincide at the origin
x = y = z = 0 at t = 0. In this sense the universe is 
homogeneous and isotropic about every such Lorentz observer
- strictly 
obeying the so called cosmological principle. This is 
manifestly obvious in co - 
moving coordinates in which the Minkowski metric describing the 
expanding ball reduces to
$$
ds^2 = dt^2 - t^2[{dr^2\over {1 + r^2}} + r^2d\Omega^2]
\eqno{(2.1)} $$
This has the form of an open FRW metric with the 
scale factor $a(t) = t$. The most appealing feature of this 
model is that at any time $t > 0$, 
every co - moving observer can see the 
entire universe : there is no horizon in the model.
$$
\int^t_0{dt\over a(t)} = \infty  ~~\forall  t > 0 \eqno{(2.2)}
$$
There is no flatness problem either as the rate of expansion 
of the universe is not constrained by any ``critical density''
parameter. This however, may be regarded as a trivial solution 
to the flatness problem. Eqn[2.1] is a solution to Einstein's
equations only if the product of the gravitational constant
and the density $G\rho$ vanishes. In canonical Einstein theory,
the Milne metric can thus only be a solution for an empty 
($\rho = 0$) universe. The model is thus put away without any
further ado. However, one can try to make out a case for a
search for models in which the universe coasts freely over 
large distances and has gravitating domains localised in
pockets. In this article we report on our study of
requirements on classes 
of ST theories in which the large scale dynamics
of a non - empty universe is described by eqn[2.1] on account 
of the vanishing of the effective long - distance gravitational
constant. 

    Consider for example, a ST theory characterised by a
non - minimal coupling of a scalar field $\phi$ with the scalar curvature
$R$, through an arbitrary function $U(\phi)$, in an effective action:

$$
S = \int\sqrt{-g}d^4x[U(\phi)R 
+ {1\over 2}\partial^\mu\phi\partial_\mu\phi
- V(\phi) + L_m] \eqno{(2.3)}
$$
Here   $V(\phi)$ is the
scalar effective potential and $L_m$ is the contribution from 
the rest of the [matter] fields. Throughout our 
analysis we treat $\phi$ as an effective classical 
field. Let $V(\phi)$ be
inclusive of an additive constant whose source could be the
characteristic cut - off mass scale that appears when we 
renormalise any quantum [matter] field. 
It could even be an arbitrary 
integration constant. It is this constant that manifests itself
as an arbitrary cosmological constant in the theory. 
The essential features of such a
theory are described in the Appendix A. As shown, the stress 
energy tensor for the rest of the matter fields has a vanishing 
co - variant divergence - in concordance with the equivalence principle.
We shall also find it essential to
include, in $L_m$, a coupling between the scalar field and a fermion field -
giving rise to an effective mass for the fermion 
determined by the local value of the scalar field. 

     If the model can consistently support a divergent
$U(\phi)\longrightarrow \infty$ 
at (say) $\phi = 0$,
flat space is a solution for $\phi = 0$ for an arbitrary
$V(\phi = 0)$. 
This gets rid of the cosmological constant problem. 
However, this is a trivial cure 
as the blowing up of $U(\phi)$ implies a vanishing of the
effective gravitational constant, and we all know [25]
that, but for gravitation, the additive constant in the 
effective potential has no dynamical role in physics.
What we want is a cure to the cosmological constant
problem in the presence of gravitation.

	We look for non trivial solutions with 
$U(\phi\longrightarrow 0)\longrightarrow \infty$. In flat spacetime, 
[with the $U(\phi)R$ term missing in eqn(2.3)],
fermion number conservation plays a key role in the existence 
and stability 
of non - trivial, NTS's.
Such solutions were  suggested by Lee and Holdom [18]
for a potential 
$V(\phi)$ with a degenerate minima 
and the analysis has been extended to non degenerate 
effective potentials [26]. 
The analysis easily extends to NTS's which are 3 - dimensional bounce 
solutions and can be summarised by referring to a variant of the Lee -
Wick model of interacting fermions $\psi$ and bosons $\phi$ defined
by the lagrangian:
$$
L  = {1\over 2}\partial_\mu\phi\partial^\mu\phi - V(\phi)
+ {1\over 2}[\partial_\mu\bar\psi\gamma^\mu\psi
- \bar\psi\gamma^\mu\partial_\mu\psi]
- m\bar\psi [1 - \phi/\phi_o]\psi
\eqno{(2.4)}
$$
The self interaction of the scalar field is parametrised in $V(\phi)$
as:
$$
V(\phi) = {1\over 2}m_\phi\phi^2[1 - \phi/\phi_o]^2
+ B[4(\phi/\phi_o)^3 - 3(\phi/\phi_o)^4]
\eqno{(2.5)}
$$
A large number $N$of fermions can get
trapped inside a spherical volume of radius $R_o$ if the energy of a
fermion in the interior is chosen to be 
less than its on - shell energy outside.
Apart from the surface term, we can work in the mean field
approximation with $\phi = \phi_o$ inside and $\phi = 0$ outside.
The two regions are separated by a shell of thickness $\approx 
(m_\phi)^{-1}$ and has a surface energy density $s \approx
m_\phi\phi_o^2/6$. The interior fermions may either have an energy
distribution of a degenerate fermion gas [with a chemical potential
$\mu$] [18(a)] or may just be a trapped relativistic gas with the
distribution described by a temperature T [18(b)]. The total 
energy of a NTS has contributions from: (1) the surface 
tension energy $E_s \approx sR_o^2$, (2) the energy of the fermions
$E_f \approx N^{4/3}/R$ and, (3) the volume energy 
[$E_V \approx V(\phi_{in})R^3$]. $V(\phi_{in})$ may be positive, 
negative or vanishing. For the degenerate case [$V(\phi_{in}) = 0$],
a NTS has total mass 
constrained by stability against gravitational collapse to a value
determined by the surface tension $s$. The soliton mass,
obtained by minimising the total energy, is just
$M = 12\pi sR_o^2$. The radius ought to be bounded from below by
$2GM$, giving a critical mass $M_c\leq [48\pi G^2s]^{-1}$.
For $s\approx (30 Gev)^3$, 
this mass bound is $ M_c\approx 10^{15}M_\odot$ 
and the radius lower bound is $R_c \approx 10^2$ light years. For
$s\approx (Mev)^3$ and the number of fermions $\approx 10^{75}$, the size
of the NTS is of the order of tens of kilo parsecs while it is still
away from the Schwarzschild bound. As one moves away from $V(\phi_{in}) = 0$,
the size of a stable soliton gets drastically effected. For
$V(\phi_{in}) > 0$, the critical mass determined by the onset of 
gravitational collapse is precipitously reduced and for $V(\phi_{in}) < 0$,
a stable NTS can exist only for very small fermion number.

	We shall now demonstrate the existence of bounce solutions in the 
theory described by eqn(2.3). Taking cue from the above 
flat space analysis, one needs a low effective surface tension 
$s\leq (Mev)^3$ to ensure a large Schwarzschild bound and 
be able to keep away
from it even for large enough NTS's. We shall consistently 
consider a low enough fermion number NTS so that gravitation 
is just a small 
perturbation over the 
essentially a flat spacetime analysis reported above.

	We look for a spherically symmetric, static solution 
described by the metric:
$$
ds^2 = e^{2u(r)}dt^2 - e^{2\bar v(r)}dr^2 - r^2[d\theta^2 + 
sin^2\theta d\varphi^2] \eqno{(2.6)}
$$
The configuration that is sought would have $\phi(r)$ locked to an almost
constant value inside a spherical domain and transiting to the exterior
region across a thin surface. 
As stated earlier, the scalar field 
gives mass to the fermions as prescribed in eqn(2.4). 
However, we shall consider a general effective potential
$V(\phi)$ that is bounded from below [figure 1] and 
having a zero at $\phi^o_{in} < \phi_o$. We consider a relativistic
fermi gas trapped in the sphere with the interior scalar field taking
values in an open interval containing $\phi^o_{in}$:
i.e. $\phi_{in}~ \epsilon~ (\phi^o_{in} -\mid\delta\mid , \phi_o)$. An
NTS that we look for has the scalar field held to a value $\phi_{in}$
in the interior of a sphere and makes a transition to $\phi = 0$ just
outside the sphere. The fermion 
density is expected to fall as $\phi$ falls across the surface as 
the fermion effective mass increases with decreasing $\phi$. 
To demonstrate
the existence of NTS's it is sufficient to ignore the $\phi$ - fermion
coupling and the matter stress energy 
across the boundary. Thus across the boundary, 
taking the trace of [A.1] enables us to eliminate 
the scalar curvature in [A.2]. The expressions for the scalar curvature
and the equation for $\phi(r)$ reduce to:
$$R[U - 3U'^2] =
{1\over 2}[ + e^{2\bar v}(\phi_{,r})^2
- 6e^{2\bar v}\phi_{,r}\phi_{,r} U'' 
+4V(\phi) - 6U'V']\eqno{(2.7)}
$$
$$
[1 - {3U'^2\over U}]\nabla^2\phi 
- (\phi_{,r})^2e^{2\bar v}{U'\over U}[{1\over 2} - 3U''] 
+ V' - {2U'V\over U} = 0 \eqno{(2.8)}
$$
Here 
$$
\nabla^2\phi = - (e^{2\bar v}\sqrt{-g}\phi_{,r})_{,r}/\sqrt{-g}
\eqno{(2.9)}
$$
Consider $W(\phi)$ and $F(\phi)$ defined by 
$$
W'(1 - {3U'^2\over U})\equiv  V'- {2U'\over U}V \eqno{(2.10a)}
$$
$$
F(U)\equiv (1 - {3U'^2\over U})^{-1}({1\over 2} - 3U''){U'\over U} 
\eqno{(2.10b)}
$$
$W$ has a minimum at $\phi = 0$ for arbitrary (but bounded) $V(0)$, 
on account 
of the diverging $U(\phi\longrightarrow 0)$. For divergent $U$, $U'$,
we choose $U$ so that $(1 - 3U'^2/U)$ does not change its (negative)
sign in the domain $(0, \phi_o)$. It would merely require $U'/U$
to be sufficiently large in this domain. 
For the same reason, it easy to
ensure $F$ to be negative in the same domain, for $W$ to have
a profile as outlined in Fig. 2, and finally, for the satisfaction of
the following
$sufficient$ condition for  
the existence of NTS's:
$$
\int^\phi_\epsilon F[U(\phi)]d\phi \longrightarrow \infty~~ iff~~
\epsilon \longrightarrow 0 \eqno{(2.11)}
$$
As long as one stays away from 
the Schwarzschild bound, spacetime can be considered to be flat to
a good approximation. In this flat - space limit,
the wave eqn(2.9) for the scalar field reduces to:
$$
{d^2\phi\over dr^2} + {2\over r}{d\phi\over dr} + F(U)[{d\phi\over dr}]^2
=  W' \eqno{(2.12)}
$$
This is an equation of a particle with $position~\phi$ at $time ~r$
moving in a potential $- W$. The second and the third terms are time
and velocity dependent damping terms. The proof of existence can be 
constructed along the lines of ``overshoot - undershoot''arguments
used by Coleman [27]. For a sufficiently large
$U'/U$, $W$ is sufficiently flat. In the interior of a soliton of
large radius $R_o$, [large enough so that the second term is
negligible in our analysis], the scalar
field is held to a constant value dependent on the fermion density
parameter $S$ [eqn(B.13)]. At $R_o$, the field falls rapidly with a
simultaneous precipitous fall of $S$. For large gradient of the scalar
field, just outside $R_o$, 
the dominant term in eqn(2.12) is the third term. The 
divergence of $U$ at $\phi = 0$ and eqn(2.11) ensure that the scalar
field would go all the way to $\phi = 0$ for an initial arbitrarily 
large scalar field gradient [$\phi '_{in}$] as
$$
{d\phi\over dr} = 
({d\phi\over dr})_oexp[\int^{\phi_{in}}_\phi F[U(\phi)]d\phi]\eqno{(2.13)}
$$
This demonstrates sufficient and not  necessary
conditions for NTS's though we have not found 
an argument that could  demonstrate their existence in general.

       For a given conserved fermion number and surface term
of an NTS, its  existence is assured by eqn(2.11) provided the initial
gradient of the scalar field at $R_o$ is large enough. Its size is 
constrained by stability. As shown in the flat space case, the larger
a NTS is, the closer should the interior potential be to zero.
Thus all large NTS's have $\phi$ approaching $\phi^o_{in}$ 
in their interiors. The 
effective gravitational constant inside all large NTS's dynamically
approach $[U(\phi^o_{in})]^{-1}$. The effective cosmological 
constant goes to zero inside large NTS's and is identically zero
outside as $U(\phi\longrightarrow 0)$ diverges.

     Similar constructions are possible in more complicated
NMC's. Consider for example a particular two component
ST theory invariant under SO(2) rotations in the
internal $\phi_1, \phi_2$ space. The action being described by
eqn[2.3] with $U(\mid\phi\mid)$ and $V(\mid\phi\mid)$ being 
functions of $\phi_1, \phi_2$ through the SO(2) invariant
$\mid\phi\mid \equiv \sqrt{\phi_1^2 + \phi_2^2}$.

       The invariance of the theory under SO(2) rotations in
$[\phi_1, \phi_2]$ space implies the conservation of any 
non - topological charge for a configuration having a compact 
support on any spacelike hypersurface $\Sigma$. The conserved
current and
the consequent conserved charge are given by:
$$
J_\mu = \phi_1\partial_\mu\phi_2 - \phi_2\partial_\mu\phi_1
\eqno{(2.14)}
$$
$$
Q = \int_\Sigma d\Sigma J^o \eqno{(2.15)}
$$

In flat spacetime, charge 
conservation plays a key role in the existence and stability 
of non - trivial, NTS's.
These are  ``Q - balls'' suggested by Coleman [27].
For these solutions we have $\phi = \phi(r) = \phi_{in}$ 
for radial coordinate  r less than 
some radius $R_o$, and $\phi$ quickly 
going to zero outside this radius. The two 
regions are separated by a transition zone of thickness independent
of the
total charge $Q$. The total charge and the internal energy of the
soliton 
are degenerate for a given volume. The surface energy is
proportional to the 
surface area of the solution. Thus the total energy for a given
charge 
of the solution is minimum for a sphere. There 
is no limit to the size of these 
solutions. The size is determined, however, 
by the total charge of the solution. Taking the cue from the 
flat space solutions we
look for 
time dependent solutions in which the scalar field rotates in the
internal
$\phi_1, \phi_2$ space with angular frequency $\omega$
$$
\phi_1 = \phi(r)sin(\omega t); \phi_2 = \phi(r)cos(\omega
t)\eqno{(2.16)}
$$
For a spherically symmetric, static solution 
described by the metric:
$$
ds^2 = g_{oo}(r)dt^2 - g_{rr}(r)dr^2 - r^2[d\theta^2 + 
sin^2\theta d\phi^2] \eqno{(2.17)}
$$
the configuration that is sought would have $\phi(r)$ locked to an
almost
constant value inside a spherical domain and transiting to the
exterior
region across a thin surface. It is straightforward to generalise [A.1] 
for this SO(2) invariant theory. The expressions for the scalar
curvature
and the equation for $\phi(r)$ reduce to:

$$R[U - 3U'^2] =$$

$$
{1\over 2}[T_{m\alpha}^\alpha - \omega^2g^{oo}\phi^2 +
g^{rr}(\phi_{,r})^2
- 6g^{rr}\phi_{,r}\phi_{,r} U'' 
+ 6g^{oo}\omega^2\phi U'
+4V(\phi) - 6U'V']\eqno{(2.18)}
$$
$$
[1 - {3U'^2\over U}]\nabla^2\phi - \omega^2\phi g^{oo}[1 -
{U'\phi\over 2U}] + V'
$$
$$ 
- {U'\over {2U}}T_{m\alpha}^\alpha  
- (\phi_{,r})^2g^{rr}{U'\over U}[{1\over 2} - 3U''] 
- {2U'V\over U} = 0 \eqno{(2.19)}
$$
Here 
$$\nabla^2\phi = - (g^{rr}\sqrt{-g}\phi_{,r})_{,r}/\sqrt{-g}
\eqno{(2.20)}$$
With $W$ and $F$ defined as before, 
the flat - space limit for
the wave eqn(2.19) of the scalar field reduces to:
$$
{d^2\phi\over dr^2} + {2\over r}{d\phi\over dr} + F(U){d\phi\over
dr}^2
= {\hat W}' \eqno{(2.21)}
$$
where 
$$
{\hat W}'\equiv W' + (1 - {3U'^2\over U})^{-1}
[- \omega^2\phi(1 - {U'\over 2U})] \eqno{(2.22)}
$$
The existence of NTS's can be demonstrated as before. 
Interestingly, the interior 
metric has an exact solution:  Plugging the metric (2.17) into
the field equation (A.1) for a uniform scalar field in the 
interior gives the following regular solution: 
$$
ds^2 = {\omega^2\phi_{in}^2\over V(\phi_{in})}dt^2 
- {dr^2\over {1 - Cr^2}} - r^2[d\theta^2 + 
sin^2\theta d\phi^2]
\eqno{(2.23)}
$$
with $C \equiv V(\phi_{in})/4U(\phi_{in})$. Expressions for the 
conserved charge and the 
internal energy of such a solution are:
$$
Q = \omega\phi_{in}^2g^{oo}v = vV/\omega \eqno{(2.24)}
$$
$$
E = 2vV \eqno{(2.25)}
$$
Here $v$ is the invariant three volume of sphere of radius $r$. 
To get the total energy one would also add the surface term which 
would be proportional to the area of the sphere. 
We have considered the NTS to be spherically symmetric. Aside from
the surface term, the total energy of a NTS for a given 
charge is degenerate in volume. The assumption of 
spherical symmetry therefore requires the contribution to the
surface energy to be positive definite. The total energy would then
be minimised for a sphere and be consistent with the assumption of
spherical symmetry.

       Once a NTS is formed with the scalar field arrested
at a value $\phi_{in}$ in the interior, the configuration
would evolve to a state of lower energy by 
changing $\phi_{in}$. The 
energy could be transmitted to a changing energy of
rotation of the scalar field in internal space and also
to the expansion of the surface wall. A drift of the scalar 
effective potential $V(\phi)$ to a vanishingly small
value while conserving the total charge must be 
accompanied by the volume $v$ blowing up to infinity.
This is the only way one could preserve the metric 
signature as  $V(\phi) \longrightarrow 0$ while conserving 
the charge. [For a bounded $g_{oo}$ as $V\longrightarrow 0$, 
$\omega \longrightarrow \sqrt{V}$. For conserved Q, this means
$v\longrightarrow \infty$ as $V\longrightarrow 0$]. The larger 
the NTS becomes, the longer would be the time expected to
synchronise the rotation of the field throughout the 
interior and the slower would be any further drift to
$V(\phi) \longrightarrow 0$. Thus a NTS would approach
$V(\phi) \longrightarrow 0$ in infinite time. 
 
     Thus in the quasi - static
approximation described above, the solution would approach
a flat configuration in the interior. The effective 
gravitational constant would approach a value 
$[U(\phi_o)]^{-1}$ irrespective of the initial value that
the NTS may be borne with. As before, 
this would dynamically generate
the universality of the induced gravitational constant and the
vanishing of the effective cosmological constant.

\vskip 1cm

\centerline{\bf Section III}

      A ``toy'' cosmology based on the Milne model [eqn(2.1)]
has characteristic features: (i) With the absence of the flatness
[fine tuning] and the horizon large scale homogeneity and isotropy
can be dynamically generated. (ii) The standard classical 
cosmological tests, viz.: the number count, angular diameter and
the luminosity distance variation with redshift are comfortably
consistent in such a cosmology. Kolb [16] has 
demonstrated concordance of classical cosmological tests 
with a coasting cosmology and we [16]
have extended his analysis to the Milne model outlined above. The 
first two tests are quite sensitive to models of galactic evolution
and for this reason have (of late) fallen into disfavour as reliable
indicators of a viable model. However the magnitude - redshift 
measurements on SN 1A have a great degree of concordance with
$\Omega_\Lambda = \Omega_M = 0$ [10]. (iii) With the scale factor 
evolving linearly with time, the Hubble 
parameter is precisely the inverse of the age t. Thus the age
of the universe
inferred from a measurement of the Hubble parameter is 1.5 times
the age inferred by the same measurement in standard matter
dominated model. Such a cosmology promises consistency 
with an older universe.  (iv) The deceleration 
parameter is predicted to vanish.

       A linear evolution of the scale factor would radically effect
nucleosynthesis in the early universe. Surprisingly, one can still
expect the following scenario to go through [28]. 
Energy conservation, in a 
period period where the baryon entropy ratio does not change, enables 
the distribution of photons to be described 
by an effective temperature $T$ that scales
as $a(t)T = $ constant [29]. 
With the age of the universe $\approx 10^{10}$
years, and $T \approx 2.7K$, one concludes that the age of the
universe at $T \approx 10^{10}K$ would be of the order of 
years [rather than seconds as in standard cosmology]. The universe would
take some $10^3$ years to cool to $10^7K$. With such a low rate of 
expansion, weak interactions remain in equilibrium for
temperatures as low as $10^8K$. The onset of nucleosynthesis is 
determined by the temperature at which deuterium burning into
heavier elements becomes a more efficient mode for destruction of 
neutrons than neutron decay. This temperature 
is completely determined by 
the baryon entropy ratio and is around $10^9K$ for interesting
values. With weak interactions still in equilibrium at this 
temperature, the neutron - proton ratio keeps falling as 
$n/p \approx exp[-15/T_9]$. There would
hardly be any neutrons left when nucleosynthesis commences at
$T_9 \approx 1$. 
However, as weak interactions are still in equilibrium, once
nucleosynthesis commences, inverse beta decay would convert 
protons into neutrons and pump them into the nucleosynthesis channel.
It turns out [28] 
that for baryon entropy ratio $10^{-8}$, there would just
be enough neutrons produced, after nucleosynthesis 
commences, to give $\approx 23.9\% ~~^4He$ and 
metalicity some $10^8$ times the metalicity produced in the early
universe in the standard scenario. This metalicity is 
of the same order
of magnitude as seen in lowest metalicity objects. The bad news is
that the residual deuterium that we get is rather low. We are
exploring the possibility of having a non uniform baryon entropy 
ratio and spallation of $^4He$ deficient baryons onto a $^4He$ rich
clouds as a mechanism to get the right deuterium while maintaining a 
low value for Lithium and other light elements.

      A Milne model within the framework of a divergent NMC would 
have a vanishing cosmological constant. Large enough gravitating
NTS domains would require a conserved fermion number in the Lee -
Wick construction reported here. A value $N_f \approx 10^{75}$
for $s \approx (Mev)^3$ can give a NTS of a size of tens of
kilo parsecs. Such an $N_f$ is of the same order as the relic 
background neutrons / photons in the universe. Thus a fermion 
species that decouples very early in the universe and which has a 
Higgs Coupling as described in eqn(2.4), would be sufficient
to provide $N_f$ for gravitating domains as large as a Halo of
a typical large structure (galaxy / local group etc.).

        How would such large NTS domains arise in the first place ?
The lack of a definitive answer to this question is the reason for
referring to the idea as a ``toy'' model. Small NTS's could form in
the early universe as a condensate of the scalar field evolves
in the universe. A large number of expanding $\phi = 0$ pockets
can constrain the condensate to colliding walls where the NTS's
would be pinched off as percolation of the entire condensate 
would be halted by the energetics of the fermi - higgs coupling.
Larger NTS's would form by collisions of smaller configurations.
Such a possibility would have a characteristic 
signature on the microwave 
background [CMB] anisotropy. A distribution of small NTS's at
$z \approx 1100$ (the surface of last scattering), that would later 
form a large NTS, would appear hotter than the background. The
number of such distributions that would be picked up in a typical
detector's beam-width should correspond to the number of large
structures at the present epoch in a scale corresponding to the 
beam width size. The rms fluctuation in temperature,  
$<\Delta T/T>$, would be determined by the inverse root of the 
average number of structures picked up by the beam width. The
variation of the rms $<\Delta T/T>$ 
with the beam width size would be the characteristic 
feature of such a model [30].

        Constraints on a NTS from lensing can be restricted to
the form of the metric [eqns(B.2, B.16)]
just inside the boundary ($R_o$)[31]. 
A metric of the form:
$$
ds^2 = dt^2 - dr^2 - r^2[d\theta^2 + 
sin^2\theta d\varphi^2]~~~~ r > R_o \eqno{(3.1)}
$$
and
$$
ds^2 \approx e^{2u_o}dt^2 - dr^2 - r^2[d\theta^2 + 
sin^2\theta d\varphi^2]~~~~ r < R_o \eqno{(3.2)}
$$
can be used to constrain $u_0$ from lensing data. It can be seen 
that a value $u_o \approx 10^{-3}$ is consistent for most lensing
objects that we have considered. Such a value could be put to
good use, as far nucleosynthesis is concerned, as it enables a
baryon at rest outside the ball to acquire a kinetic energy
$\approx Mev$ as it goes inside the ball. In 
an inhomogeneous model - having a different baryon entropy ratio 
in the interior and exterior of NTS's, one could easily have 
a $^4He$, metal rich, interior, and a proton rich exterior, by
the time the universe cools to temperatures below $\approx 10^7K$.
At lower temperatures, the proton rich clouds in the exterior 
can spall over the interior matter and lead to deuterium formation
[28].

\centerline{\bf IV. Conclusion} 
	 
	 There are two distinct aspects of the work presented
here. First is the viability, advantages and consequences of a 
coasting cosmology.   The second aspect is the possible
realisation of such a coasting. The absence of the horizon, 
flatness and age problems distinguish a coasting cosmology.

	  We have demonstrated that in a whole class of 
scalar tensor theories in which the
non - minimal coupling diverges and for which the classical 
effective potential vanishes at some point, classical scalar
field condensates can occur as NTS's.  The effective 
gravitational constant inside all large domains would approach a 
universal value and the effective cosmological constant
would drift to zero. The dynamical tuning of the effective
cosmological constant to a small value and the effective
gravitational constant to a universal value are compelling
features - enough to explore the possibility of  raising the toy
model described here to the status of a viable cosmology.

      The cosmology described in this article essentially
requires the scale factor to coast linearly with time. The
particular model outlined here is deficient in several respects.
Firstly we have treated the scalar field as purely classical.
The stability of the NTS against decay has not been studied. We
feel that the peculiar behaviour of the NMC at $\phi = 0$
would prevent any stable $\phi$ particle states to materialise
near $\phi = 0$.

     The idea of exploring NMC for the purpose of getting
stress energy of the scalar
field condensate to compensate
the cosmological constant was also suggested by Dolgov[4]. It
was shown however that the NMC itself diverged. It was indeed
suggested that a spatial variation of the scalar field -
and hence a gravitational constant be explored for a non -
trivial model. What we have shown is that in a model where the
NMC diverges over most of the space and is finite over
compact domains, the compact domains can be expected to
inflate. Conservation of charge and energy would then ensure that
the effective cosmological constant approaches zero inside these
domains and identically vanishes outside. Instead of working
with a NMC that diverges as a function of $\phi$ as
$\phi \longrightarrow 0$, one could equally well work with 
$U(\phi) = \epsilon\phi^2$ in a model where 
$V(\phi \longrightarrow \infty) = $ constant. In such a model, 
as discovered by Dolgov [23], there are solutions that has the
scalar field rising to infinity as a function of time. If
$V(\phi)$ has a zero then, with a judicious Higgs coupling 
of a fermion as described in this article, one would get 
results similar to those obtained.

\vskip 1cm
\centerline{Acknowledgements:}

	  This work has benefitted greatly from
discussions with Profs. T. W. Kibble, G. W. Gibbons and 
T. Padmanabhan. 

\vskip 1cm

\centerline{\bf Appendix A}
     
	We describe properties of the 
scalar tensor theory and derive expressions for a conserved
pseudo energy tensor for the theory. The theory is described 
by the action:

$$
S = \int\sqrt{-g}d^4x[U(\phi)R + L_\phi + L_m] \equiv 
\int\sqrt{-g}d^4x[U(\phi)R 
+ {1\over 2}\partial^\mu\phi\partial_\mu\phi
- V(\phi) + L_{m}] \eqno{(2.3)}
$$
Here  $\phi$ is the scalar field,
$U(\phi)R$ is a 
non - minimal coupling of the scalar field
with the scalar curvature, and $V(\phi)$ is the
scalar effective potential. $L_m$ consists of a fermion field
$\psi$, together with its Higgs coupling with $\phi$, and $L_w$
(the rest of the matter lagrangian) independent of $\phi$ and
$\psi$:
$$
L_m \equiv L_\psi + L_{\psi,\phi} + L_w 
\equiv {1\over 2}[\bar\psi\overleftarrow{D}_\mu\gamma^\mu\psi
- \bar\psi\gamma^\mu\overrightarrow{D}_\mu\psi]
- m(1 - {\phi\over \phi_o})\bar\psi\psi + L_w
$$
Here $D_\mu$ is the spin covariant derivative [see eg. [33]]
$$
\overrightarrow{D}_\mu\psi = (\partial_\mu + \Gamma_\mu)\psi
$$
$$
\bar\psi\overleftarrow{D}_\mu = (\partial_\mu\bar\psi - \bar\psi\Gamma_\mu)
$$
$\Gamma_\mu$ are the spin connection [Fock - Ivanenko] coefficients
defined by:
$$
D_\nu\gamma_\mu \equiv \partial_\nu\gamma_\mu 
- \Gamma^\alpha_{\mu\nu}\gamma_\alpha + [\Gamma_\nu,\gamma_\mu]
$$
Requiring the action to be 
stationary under variations of the metric tensor and the fields 
$\phi, \psi$, gives the
equations of motion:
$$
U(\phi)[R^{\mu\nu} - {1\over 2}g^{\mu\nu}R] = -{1\over 2}[T_w^{\mu\nu} 
+ T_\phi^{\mu\nu} + T_{\phi,\psi}^{\mu\nu} + T_\psi^{\mu\nu}   
+ 2U(\phi)^{;\mu;\nu} 
- 2g^{\mu\nu}U(\phi)]^{;\lambda}_{;\lambda}] \eqno{(A.1)}
$$
$$
g^{\mu\nu}\phi{;\mu;\nu} + {\partial V\over {\partial\phi}} 
- R{\partial U \over {\partial \phi}} - {m\over \phi_o}\bar\psi\psi  
= 0 \eqno{(A.2(a))}
$$
$$
\gamma^\mu D_\mu\psi + m(1 - {\phi\over \phi_o})\psi = 0\eqno{(A.2(b))}
$$
$$
D_\mu\bar\psi\gamma^\mu - m(1 - {\phi\over \phi_o})\bar\psi = 0\eqno{(A.2(c))}
$$
Here $T_w^{\mu\nu}$, $T_\psi^{\mu\nu}$and $T_{\phi,\psi}^{\mu\nu}$
are the energy momentum tensors constructed from
$L_w$ and $L_\psi + L_{\psi,\phi}$ respectively, and
$$
T_\phi^{\mu\nu} = \partial^\mu\phi\partial^\nu\phi 
- g^{\mu\nu}[{1\over 2}\partial^\lambda\phi\partial_\lambda\phi -V(\phi)]
\eqno{(A.3)}
$$
$L_w$ is  independent of $\phi$. To
examine this theory viz - a - viz the equivalence principle, we have
to explore conditions under which $T^{\mu\nu}_{;\nu} = 0$.
Eqns.(2(a), 2(b)) show that the portion of lagrangian 
$L_\psi + L_{\psi,\phi}$ is null. The stress tensor is given by 
the following generalisation of the familiar flat spacetime
expression [see eg. [34]]:
$$
\Theta^\mu_\nu \equiv T_{\psi\nu}^\mu + T_{\psi,\phi,\nu}^\mu
= -{1\over 2}[\bar\psi\overleftarrow{D}_\nu\gamma^\mu\psi
- \bar\psi\gamma^\mu\overrightarrow{D}_\nu\psi]\eqno{(A.4)}
$$
When applied to any spinor or any ``spin - matrix'' such as the
Dirac matrices, one replaces the ordinary derivative by the spin -
covariant derivative [33]. The covariant divergence of 
[A.4] is easily seen to reduce to:
$$
\Theta^\mu_{\nu;\mu} = {m\over \phi_o}\partial_\nu\phi\bar\psi\psi
\eqno{(A.5)}
$$
Thus there is a violation of equivalence principal as far as the 
fermi field is concerned. However, in a region where the scalar
field gradient, $\partial_\mu\phi$, vanishes, the covariant divergence
of the fermion field stress tensor vanishes. For the rest of the
matter fields, the equivalence principal holds strictly, i.e.:
$T^{\mu\nu}_{w;\nu} = 0$. To see this, consider the covariant 
divergence of [A.1]. From the contracted 
Bianchi identity satisfied by the Einstein tensor, we get
$$
U(\phi)_{,\nu}[R^{\mu\nu} - {1\over 2}g^{\mu\nu}R] = 
-{1\over 2}[T^{\mu\nu}_{w;\nu} + t^{\mu\nu}_{;\nu}
+ \Theta^{\mu\nu}_{;\nu}] \eqno{(A.6)}
$$
with 
$$t^{\mu\nu} \equiv T_\phi^{\mu\nu} +  2U(\phi)^{;\mu;\nu} -
2g^{\mu\nu}U(\phi)^{;\lambda}_{;\lambda} \eqno{(A.7)}$$
Using the identity: 
$U(\phi)^{;\rho}R_{\rho\alpha} = U(\phi)^{;\lambda}_{;\lambda ;\alpha}
- U(\phi)^{;\lambda}_{;\alpha;\lambda}$ 
and the eqn(A.5), this reduces to
$$
-{1\over 2}U(\phi)^{,\mu}R] = 
-{1\over 2}[T^{\mu\nu}_{w;\nu} + T^{\mu\nu}_{\phi;\nu}
+ \partial^\mu({m\over \phi_o})\bar\psi\psi] 
$$
Finally, using the equation of motion for the scalar field [A.2a],
all the $\phi$ dependent terms cancel the left hand side -
giving the vanishing of the covariant divergence of the (w-) matter
stress energy tensor.
 
    One can find the expression for a conserved pseudo energy 
momentum tensor that would be conserved. To achieve this we proceed to 
express the vanishing covariant divergence of the matter stress energy 
tensor as:
$$
[\sqrt{-g}T^\nu_{m\mu}]_{,\nu} 
- {1\over 2}g_{\tau\beta,\mu} \sqrt{-g}T^{\tau\beta}_m = 0 \eqno{(A.8)}
$$
To cast the LHS of the above equation into a total ordinary divergence one 
has to seek a representation of the second quantity in terms of an
ordinary total divergence. This can be done as follows. 
First we make use of the
equation of motion (A.1) 
to express the matter stress energy tensor in terms of the 
other fields and the metric -dependent quantities:
$T_w^{\tau\beta} \equiv -t^{\tau\beta} - \Theta^{\tau\beta} 
- 2U(\phi)G^{\tau\beta}$. Second, note 
that the right hand side of this expression is merely the variational
derivative of 
$$ 
J \equiv 2\int\sqrt{-g}d^4x[U(\phi)R + L_\phi 
+ L_\psi + L_{\psi,\phi}] \eqno{(A.9)}
$$
under variations of the metric tensor, with boundary conditions that require
the vanishing of metric and its first derivative variations on 
the boundary of a (3+1) - dimensional manifold over 
which this  integral has been taken.
Consider the standard decomposition of 
$\sqrt{-g}R$ into a pure divergence term and a simple expression involving 
only the metric and its first derivatives:
$$
\sqrt{-g}R = {\rm A}
+ [\sqrt{-g}g^{\sigma\rho}\Gamma^\alpha_{\sigma\alpha}]_{,\rho}
-  [\sqrt{-g}g^{\sigma\rho}\Gamma^\alpha_{\sigma\rho}]_{,\alpha}\eqno{(A.10)}
$$
with 
$$
{\rm A} \equiv
\sqrt{-g}g^{\sigma\rho}[\Gamma^\alpha_{\sigma\rho}]\Gamma^\beta_{\alpha\beta}
- \Gamma^\alpha_{\beta\rho}]\Gamma^\beta_{\alpha\sigma}]\eqno{(A.11)}
$$
It follows that the functional derivative of $J$ with respect to the 
metric tensor is the same as that of  
$$
H \equiv \int d^4x [{\rm B} + \sqrt{-g}(L_\phi
+ L_\psi + L_{\psi,\phi})]\eqno{(A.12)}
$$
where
$$
{\rm B} \equiv [U{\rm A} 
- \sqrt{-g}g^{\sigma\rho}\Gamma^\alpha_{\sigma\alpha}U_{,\rho}
+ \sqrt{-g}g^{\sigma\rho}\Gamma^\alpha_{\sigma\rho}U_{,\alpha}]\eqno{(A.13)}
$$
In other words
$$
\sqrt{-g}UG_{\mu\nu} + 
\sqrt{-g}[U_{;\mu;\nu} - g_{\mu\nu}U^{;\alpha}_{;\alpha}]
+ {1\over 2}\sqrt{-g}T_{(\phi + \psi)\mu\nu}
$$
$$= {\partial\over \partial g^{\mu\nu}}[{\rm B} + \sqrt{-g}L_{\phi + \psi}]
- [{{\partial({\rm B} + \sqrt{-g}L_{\phi + \psi})}\over 
{\partial g^{\mu\nu}_{,\lambda}}}]_{,\lambda}\eqno{(A.14)}
$$
This is just a generalisation of the standard procedure in GTR
[32].
Definining $\hat {\rm B} \equiv {\rm B} + \sqrt{-g}L_{\phi + \psi}$, 
the expression for the 
ordinary derivative of $\hat {\rm B}$ and 
the field equation for the fields
$\phi, \psi$  easily enable us to express the 
second term in eqn(A.8) as a total
divergence. This gives
$$
[\sqrt{-g}T^\nu_{m\mu} - \hat {\rm B} \delta^\nu_\mu
- {{\partial\hat{\rm B}}\over {\partial g^{\tau\beta}_{,\nu}}}g^{\tau\beta}_{,\mu}
-{{\partial \hat{\rm B}}\over {\partial\phi_{,\nu}}}\phi_{,\mu}
-{{\partial \hat{\rm B}}\over{\partial\psi_{,\nu}}}\psi_{,\mu}
-\bar\psi_{,\mu}{{\partial \hat{\rm B}}\over{\partial\bar\psi_{,\nu}}}]_{,\nu}
= 0
\eqno{(A.15)}
$$
For $\nu = o$  the expression within the brackets integrated over
a spacelike hypersurface is thus invariant under time translations for 
a distribution of matter and the rest of the terms in  (A.15) having
a compact support over the surface. 
This is the expression for the pseudo energy momentum tensor that we seek.
The quantity
$$
P_\mu \equiv \int_\Sigma d\Sigma[\sqrt{-g}T^o_{w\mu} 
- \hat {\rm B} \delta^o_\mu
- {{\partial\hat{\rm B}}\over {\partial g^{\tau\beta}_{,o}}}g^{\tau\beta}_{,\mu}
-{{\partial \hat{\rm B}}\over {\partial\phi_{,o}}}\phi_{,\mu}
-\bar\psi_{,\mu}{{\partial \hat{\rm B}}\over {\partial\bar\psi_{,o}}}
-{{\partial \hat{\rm B}}\over {\partial\psi_{,o}}}\psi_{,\mu}
]\eqno{(A.16)}
$$
evaluated on a constant spacelike hypersurface $\Sigma$, is thus conserved. 
This may be viewed as the generalisation of the energy momentum four vector
for the scalar - tensor theory described by eqn[2.3]. The 
formalism presented here is general and can be used to determine
the energy momentum four vector for any Brans - Dicke theory in 
particular.
As in standard general relativity, $P_\mu$
is not a generally covariant four vector as ${\rm A}$ and ${\rm B}$ are not 
scalar densities. The intrinsic non - covariance of the energy momentum 
density of the gravitational field has its origin in the intimate 
connection between geometry and the gravitational field. Had the expression
been covariant, one could always have gone into a preferred [freely - falling] 
frame to ensure vanishing of an arbitrary localised gravitational field.

        This expression for the energy is sufficient for the present article. 
For a $U(\phi)$ that is well behaved [bounded], it is possible to
reduce the energy as an integral over a two sphere. This is not 
case for the present article [$U(\phi)$ has been chosen to diverge].
[see Bose, Lohiya-- for the reduction].

\vskip 1cm
\centerline{\bf Appendix B}

       We derive the form for the metric of a NTS satisfying a ``weak
field approximation'' that would justify retaining only a first order
deviation from a flat metric. The metric can be expressed in terms 
of the spherical [Schwarzschild] coordinates:
$$
ds^2 = e^{2u}dt^2 - e^{2\bar v}dr^2 - r^2[d\theta^2 + 
sin^2\theta d\varphi^2] \eqno{(B.1)}
$$
or in terms of isotropic coordinates:
$$
ds^2 = e^{2u}dt^2 - e^{2v}(d\rho^2 + \rho^2d\theta^2 
+ \rho^2\sin^2\theta d\varphi^2) \eqno{(B.2)}
$$
related to each other by:
$$
r = \rho e^v \eqno{B.3}
$$
We look for a solution describing the scalar field trapped to a 
value $\phi = \phi_{in}$ in the interior of a sphere of radius
$R_o$ and making a transition across a thin surface to $\phi = 0$
outside. The fermi gas trapped inside the soliton is described by 
the familiar distribution in momentum space: $n_k$, $k$ being the 
momentum measured in an appropriate local frame that depends on 
$r$ or $\rho$. The fermion energy density is given by [18]:
$$
W = {2\over 8\pi^3}\int d^3kn_k\epsilon_k\eqno{(B.4)}
$$
with $\epsilon_k = \sqrt{k^2 + (m - f\phi_{in})^2}$. The number
density $\nu$ and the non - vanishing components of fermion stress 
energy tensor are:
$$
\nu = {2\over 8\pi^3}\int d^3kn_k\eqno{(B.5)}
$$
$$
T^t_t = W
$$
$$
T^r_r = T^\theta_\theta = T^\varphi_\varphi = T^\rho_\rho
\equiv -T = -{2\over 8\pi^3}\int d^3kn_k{k^2\over 3\epsilon_k}\eqno{(B.6)}
$$
The trace of the stress tensor is just:
$$
T^\mu_\mu = W - 3T = (m - f\phi_{in})S\eqno{(B.7)}
$$
with S the scalar density:
$$
S = {2\over 8\pi^3}\int d^3k{n_k\over \epsilon_k}(m -f\phi_{in})\eqno{(B.8)}
$$
Defining $G_{in} \equiv U(\phi_{in})^{-1}$ as the effective
interior ``gravitational constant'', the metric field equation in the 
interior can be expressed in the spherical coordinates as:
$$
r^2G^t_t = e^{-2\bar v} - 1 -  e^{-2\bar v}r{d\bar v\over dr}
= - 8\pi Gr^2[W + V(in)]\eqno{B.9}
$$
$$
r^2G^r_r = e^{-2\bar v} - 1 +  e^{-2\bar v}r{du\over dr}
= 8\pi Gr^2[T - V(in)]\eqno{B.10}
$$
$$
r^2G^\theta_\theta = e^{-2\bar v}[r^2{d^2u\over dr^2} 
+ [1 + r{du\over dr}]r{d\over dr}(u - \bar v)]
=  8\pi Gr^2[T - V(in)]\eqno{B.11}
$$
The scalar field satisfies:
$$
\phi^{;\mu}_{;\mu} + V' - fS - U'R = 0 \eqno{B.12}
$$
Taking the trace of the Einstein  tensor $G^\mu_\mu$ in 
eqns(B.9 - 11), the
Ricci scalar $R$ can be substituted in eqn(B.12). The condition for
the existence of a $\phi = \phi_{in} = $ constant for 
$0\leq r \leq R_o$ reduces to:
$$
V' - fS - 2V{U'\over U} = {U'\over U}(m - f\phi_{in})S\eqno{(B.13)}
$$
For a large enough choice for the gradient of the NMC, this 
condition can be satisfied in an open interval containing 
$\phi^o_{in}$ for any S. The form for the metric in the linear 
approximation follows from eqns(B.9 - 11). Defining 
$\hat{C} \equiv 8\pi G[W + V(\phi_{in})]$, we get 
$\bar v = - \hat{C}r^2/6$. The expression for u follows from:
$$
2e^{-\bar v}r[{du\over dr} + {d\bar v\over dr}] 
= 8\pi Gr^2[T + W] \equiv \tilde{C}r^2\eqno{(B.14)}
$$
Whence, in the linear approximation being followed,
$$
{du\over dr} + {d\bar v\over dr} 
= {1\over 2}\tilde{C}r\eqno{(B.14)}
$$
From the expression derived for $\bar v$, we get:
$$
u = u_o + {r^2\over 2}[{\tilde{C}\over 2} + {\hat{C}\over 3}]
\eqno{(B.15)}
$$
In isotropic coordinates, this transforms to 
$$
u = u_o + {\rho^2\over 2}[{\tilde{C}\over 2} + {\hat{C}\over 3}]
\eqno{(B.16)}
$$
and $v = \hat{C}\rho^2/12$. $u_o$ is a small negative constant
that determines the rate at which clocks tick at the origin 
$r = \rho = 0$. This is determined by integrating the field 
eqn(A.1) from outside the NTS, where $u_o$ vanishes, across the
surface of the NTS into the interior and all the way upto the origin.
This constant would determine the bending of a null ray
as it moves across the surface of the NTS.

\vfil\eject

\vskip 2cm

\centerline{\bf References:}
\item{1.} P. J. Steinhardt and M. S. Turner,
Phys. Rev.$\underline{D29}$,
2162 (1984)
\item{2.}G. F. Smoot et al., Astrophys. J. Lett,$\underline{396}$, L1
(1992)
\item{3.}A. Albrecht and P. J. Steinhardt, Phys. Rev. Lett.
$\underline{48}$, 1220 (1982); A. D. Linde, Phys. Lett. 
$\underline{108B}$, 389 (1982)
\item{4.}A. D. Linde, Phys. Lett.$\underline{129B}$, 177 (1983)
\item{5.}F. C. Adams and K. Freese, Phys. Rev.$\underline{D51}$,
6722 (1995)
\item{6.}F. C. Adams, K. Freese and A. H. Guth, Phys. Rev.
$\underline{D43}$, 965 (1991)
\item{7.}R. H. Brandenberger, hep-ph/9701276 (1997)
\item{8.}M. Bucher, A. S. Goldhaber and N. Turok, Phys. Rev.
$\underline{D52}$, 3314 (1995)
\item{9.} D. La and P. J. Steinhardt, Phys. Rev. Lett.
$\underline{62}$, 376 (1989)Phys. Rev. Lett.
$\underline{64}$, 2740 (1990)
\item{10.}B. P. Schmidt et al., astro-ph/ 9805200 (1998); 
P. M. Garnavich et al., astro-ph/ 9806396 (1998)
\item{11.}P. Coles and G.F.R. Ellis, ``Is the universe open or 
closed ?'' Cambridge Univ. Press (1997)
\item{12.}B. Campbell, K. Olive and A. Linde, Nuc. Phys. 
$\underline{B355}$, 146 (1991)
\item{13.}T. Damour and A. M. Polyakov, Nuc. Phys. 
$\underline{B423}$, 532 (1994); Gen. Rel. and Grav. $\underline{26}$,
1171 (1994)
\item{14.}J. G. Bellido and A. D. Linde,Phys. Rev.
$\underline{D52}$, 6730 (1995); ibid:Phys. Rev.
$\underline{D52}$, 6739 (1995); J. D. Barrow and K. Maeda, Nuc. Phys.
$\underline{B341}$, 294 (1990); A. M. Green and A. R. Liddle, Phys.
Rev. $\underline{D54}$, 2557 (1996); D. Lyth and E.D. Stewart, Phys.
Rev. $\underline{D54}$, 7186 (1996); D. F. Torres and H. Vucetich,
Phys. Rev. $\underline{D54}$, 7373 (1996); D. Langlois, Phys. Rev.
$\underline{D54}$, 2447 (1996)
\item{15.} E.A.Milne Relativity,Gravitation and World Structure
(Oxford) (1935)
\item{16.}E. Kolb, Astrophys. J. $\underline{344}$, 543 (1989); 
M. Sethi \& D. Lohiya, ``Aspects of a Coasting Universe'', Univ.
Delhi preprint, 1996 [also GR15 proceedings 1997]
\item{17.}H. Dehnen \& O. Obregon, Ast. and Sp. Sci. $\underline{17}$,
338 (1972); ibid. $\underline{14}$, 454 (1971)
\item{18.}(a) T. D. Lee, Phys. Rev. $\underline{D35}$, 3637 (1987);
T. D. Lee \& Y. Pang, phys. Rev. $\underline{D36}$, 3678 (1987);
(b) B. Holdom, Phys. Rev. $\underline{D36}$, 1000 (1987)
\item{19.}F. C. Adams et al, Phys. Rev.$\underline{D47}$,
426 (1993)
\item{20.}C. G. Callan, S. Coleman \& R. Jackiw Ann. Phys. 
$\underline{59}$, 42 (1972)
\item{21.}A.Zee,in 1981 Erice Conference Proceedings,ed. A.Zichichi,
Plenum.
\item{22.} S.Deser, Ann.Phys. $\underline{59}$, 248 (1970)
\item{23.}A. D. Dolgov, in "The very Early Universe" eds. 
G. W. Gibbons, 
S. W. Hawking and S. T. Siklos, (1982) CUP.
\item{24.} Madsen, Class.Quantum Grav. $\underline{5}$, 627 (1988)
\item{25.} S. Weinberg, Rev. Mod. Phys 1 (1989)
\item{26.} W. M. Cottingham \& R. V. Mau Phys. Rev. $\underline{D44}$,
1652 (1991)
\item{27.} S. Coleman, Nuc. Phys. $\underline{B262}$, 263 (1985)
\item{28.} A. Batra et al., ``Nucleosynthesis in a simmering
universe'' GR 15 proceedings [1997]
\item{29.} S. Weinberg, ``Gravitation and Cosmology'', J. Wiley (1972)
\item{30.} S. Gehlaut and D. Lohiya, 
``Evolution of perturbations in a coasting 
cosmology'', GR15 proceedings (1997) 
\item{31.} M. Safonova and D. Lohiya, ``Gravity balls in alternative
induced gravity model - gravitational lens effects'', GR15 proceedings
(1997)
\item{32.} See eg. R. Adler, M. Bazin \& M. Schiffer, ``Introduction
to General Relativity'' McGraw - Hill (1975) chapter 11.
\item{33.} H. Pagels, Ann. Phys. $\underline{31}$, 64 (1965)
\item{34.} See eg. C. Ikzykson and J. Zuber, ``Quantum field theory''
McGraw - Hill (1985).

\bye